\documentclass[mathleft
]{an}
\usepackage{graphicx}
\usepackage{times}
\newcommand{\oii}{\mbox{[O\,{\scriptsize II}]}}

\begin{document}

\Pagespan{1}{}
\Yearpublication{2009}%
\Yearsubmission{2009}%
\Month{11}%
\Volume{999}%
\Issue{99}%

\title{The Evolution of Spheroidal Galaxies in Different Environments}

\author{Alexander Fritz\inst{1}\fnmsep\thanks{Corresponding author:
  \email{afritz@gemini.edu}\newline} \and  Bodo L.\ Ziegler\inst{2}}
  
\titlerunning{The Evolution of Spheroidal Galaxies in Different Environments}
\authorrunning{A. Fritz \& B.\ L.\ Ziegler}
\institute{Gemini Observatory, 670 N.\ A'ohoku Place,
Hilo, HI 96720, USA
\and 
European Southern Observatory, Karl-Schwarzschild-Strasse 2, 85748
Garching bei M\"unchen, Germany}

\received{\today}


\abstract{%
We analyse the kinematic and chemical evolution of 203 distant spheroidal
(elliptical and S0) galaxies at $0.2\nobreak<\nobreak z\nobreak<\nobreak0.8$ which are located in different
environments (rich clusters, low-mass  clusters and in the field). VLT/FORS
and CAHA/MOSCA spectra with intermediate-resolution have been
acquired to measure the  internal kinematics and stellar populations of the
galaxies. From HST/ACS and WFPC2 imaging, surface brightness profiles
and structural parameters were derived for half of the galaxy sample. The
scaling relations of the Faber-Jackson relation and Kormendy relation
 as well as the Fundamental Plane indicate a moderate evolution for the
whole galaxy population in each density regime. In all environments, S0
galaxies show a faster evolution than elliptical galaxies. For the cluster
 galaxies a slight radial dependence of the evolution out to one virial radius is
found. Dividing the samples with respect to their mass, a mass dependent
evolution with a stronger evolution of lower-mass galaxies ($M<2\times10^{11} M_{\sun}$)
is detected. Evidence for recent star formation is provided by blue colours and
weak \oii\ emission or strong H$\delta$ absorption features in the spectra. The
results are consistent with a down-sizing formation scenario which is
independent from the environment of the galaxies.}

\maketitle

\section{Introduction}

Hierarchical structure formation models based on cold dark matter particles
predict different evolutionary paths and assembly time-scales for the densest
environments of clusters of galaxies and the lowest densities of isolated
galaxies (Baugh et al. 1998; Somerville \& Primack 1999). 
From typical age variations of 43\% in the overall stellar content
between cluster and field galaxies, a late mass assembly for massive spheroidal
galaxies of smaller sub-units is expected. However, the involved physical
processes (star formation rates, gas cooling and heating) remain
poorly understood and thus the simplified baryonic recipes are
interpreted and implemented differently.
In this paradigm, field early-type galaxies are assembled through merging and
accretion events of smaller clumps and fragments on longer time-scales up to
the recent past (De Lucia et al. 2004). The environment is therefore suggested
to play an important role to shape the formation and evolution of 
early-type (spheroidal) galaxies and their internal properties.

To shed more light on the formation and assembly history of spheroidal galaxies,
we have conducted an extensive observational campaign using a combination
of high signal-to-noise ($S/N$), intermediate-resolution VLT/FORS and CAHA/MOSCA
spectroscopy together with HST/ACS plus WFPC2 imaging and deep ground-based
multi-band photometry. Target galaxies are residing in environments of various
densities, from massive rich galaxy clusters over low-mass (poor) galaxy
clusters to the isolated field. The motivation of the project was to look
how the physical properties of early-type galaxies, as well as their
sub-classes of elliptical and S0 galaxies, depend on their environment.
The evolution of galaxies in luminosity, size, mass and their 
stellar populations were intercompared to test possible environmental effects 
and to constrain the theoretical formation and evolution model predictions.
Throughout this work a cosmology for a flat, low-den\-sity Universe is assumed
with $\Omega_{m}=0.3$, $\Omega_{\Lambda}=0.7$ and
$H_0=70$\,km\,s$^{-1}$\,Mpc$^{-1}$.

\section{Observational Data}

Previous spectroscopic studies at moderate redshift were limited to a small
number of the more luminous galaxies. To overcome bias and selection problems
of small samples, we focus in this work on a large number of 96 cluster
members in two massive rich galaxy clusters Abell~2218 (Ziegler et al. 2001)
and Abell~2390 (Fritz et al. 2005) and 68 cluster members in six low-mass
galaxy clusters at $0.2\nobreak<\nobreak z\nobreak<\nobreak0.3$.
(Fritz et al. 2004; Fritz et al. 2010 in prep). 
The low-mass clusters were selected to have very low X-ray luminosities
($L_{X}<4\times 10^{43}$~erg/s)\footnote{$L_{X}$ between 0.1--2.4 keV}
1.0 dex lower than X-ray bright rich CNOC clusters, and poor optical richness
class. Sphe\-roidal candidates in rich clusters were
selected based on deep $UBI$ ground-based imaging using colour-colour
criteria that effectively removes fore- and background objects. Sphe\-roidal
candidates in poor clusters were selected by a combination of
$BVRI$  photo\-metry and spectroscopic redshifts.
The galaxies span a wide range in luminosity ($M_{B}\ga M^{\ast}+2$),
corresponding for A\,2390 to $21.4<B<23.3$, and a wide field-of-view 
of $\sim 10'\times 10'$ ($1.6\ h^{-1}_{70}\times 1.6\ h^{-1}_{70}$ \newline Mpc$^{2}$).

\begin{figure}
\includegraphics[width=0.24\textwidth]{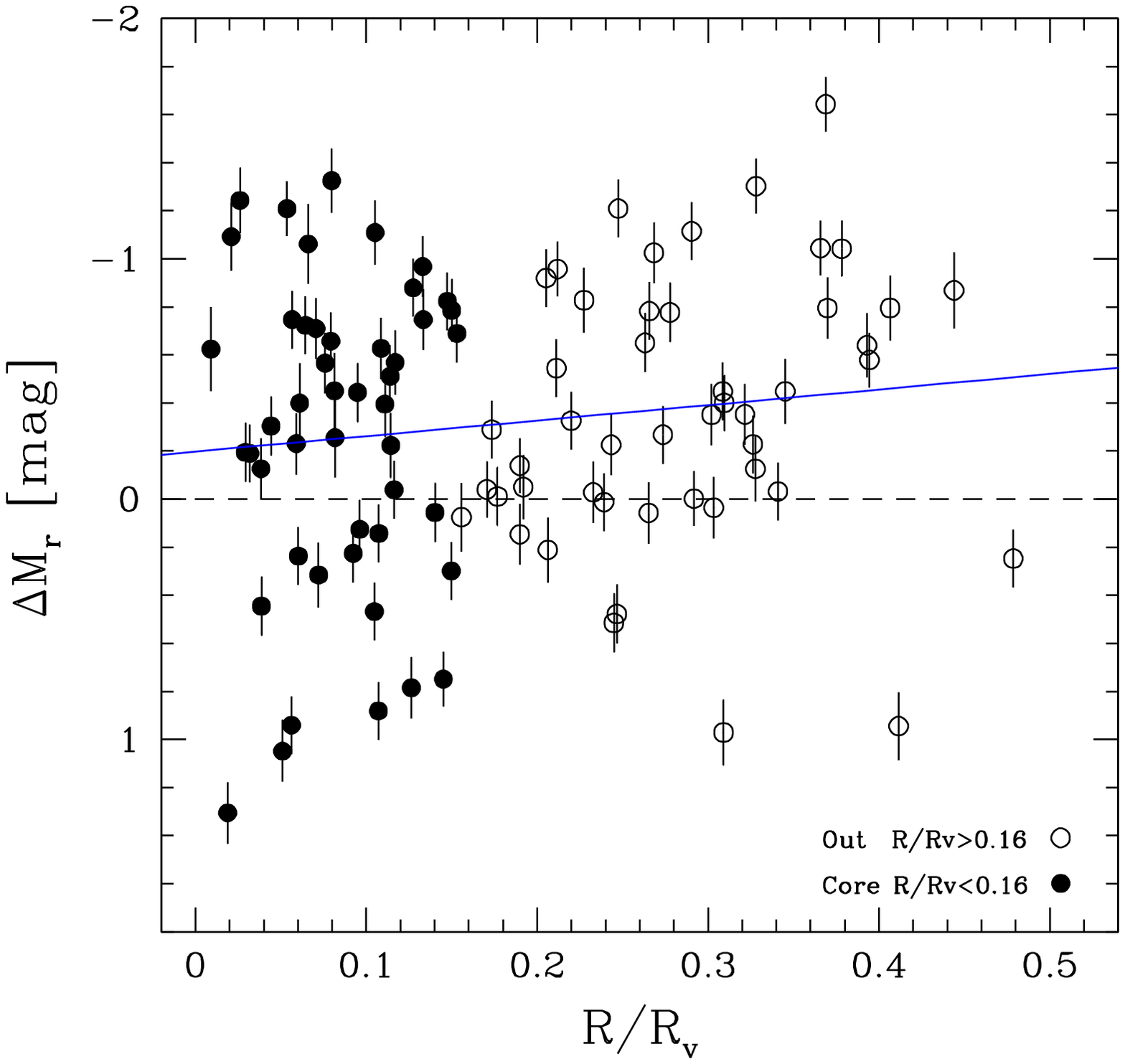}
\includegraphics[width=0.24\textwidth]{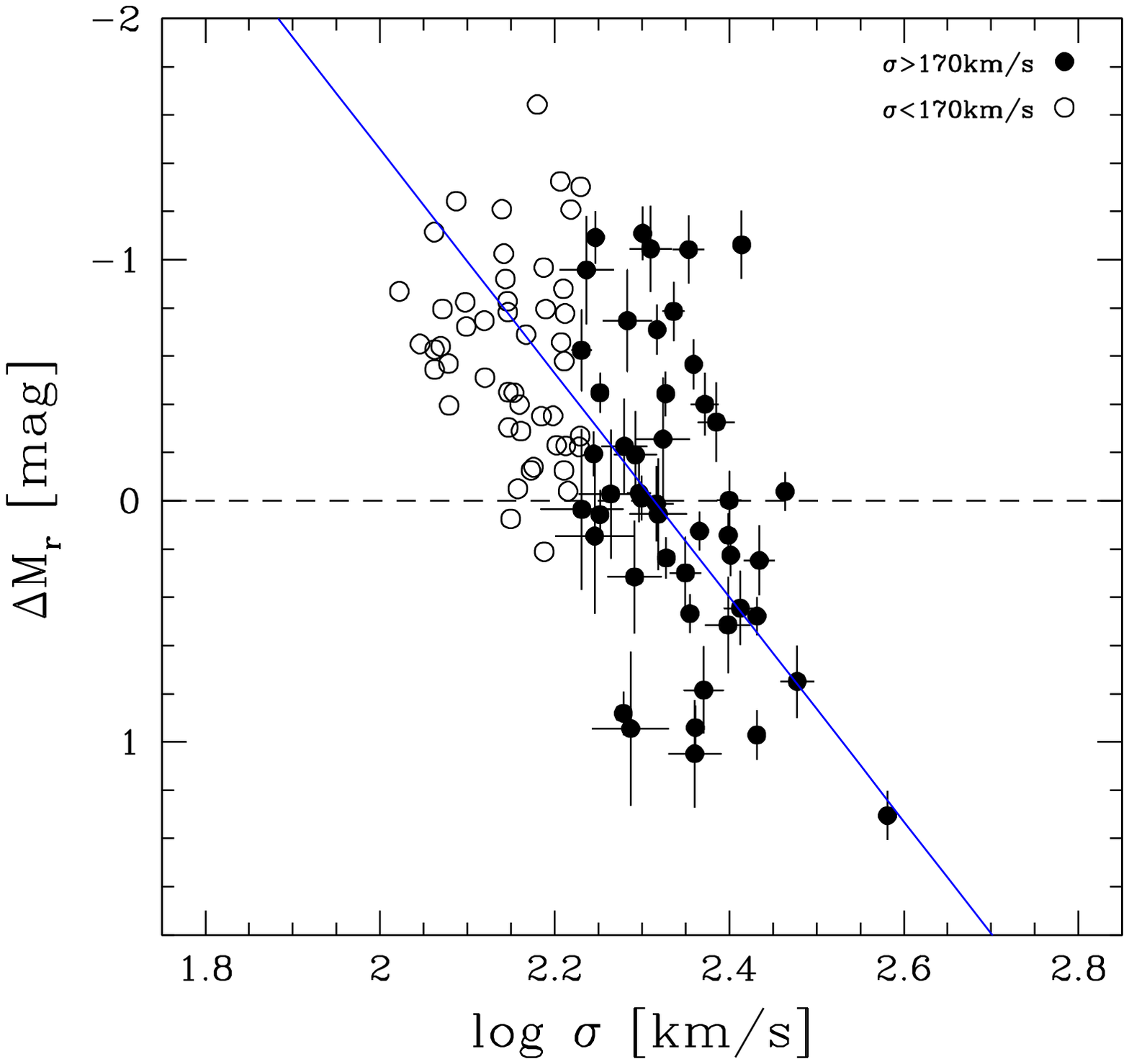} \\
\includegraphics[width=0.24\textwidth]{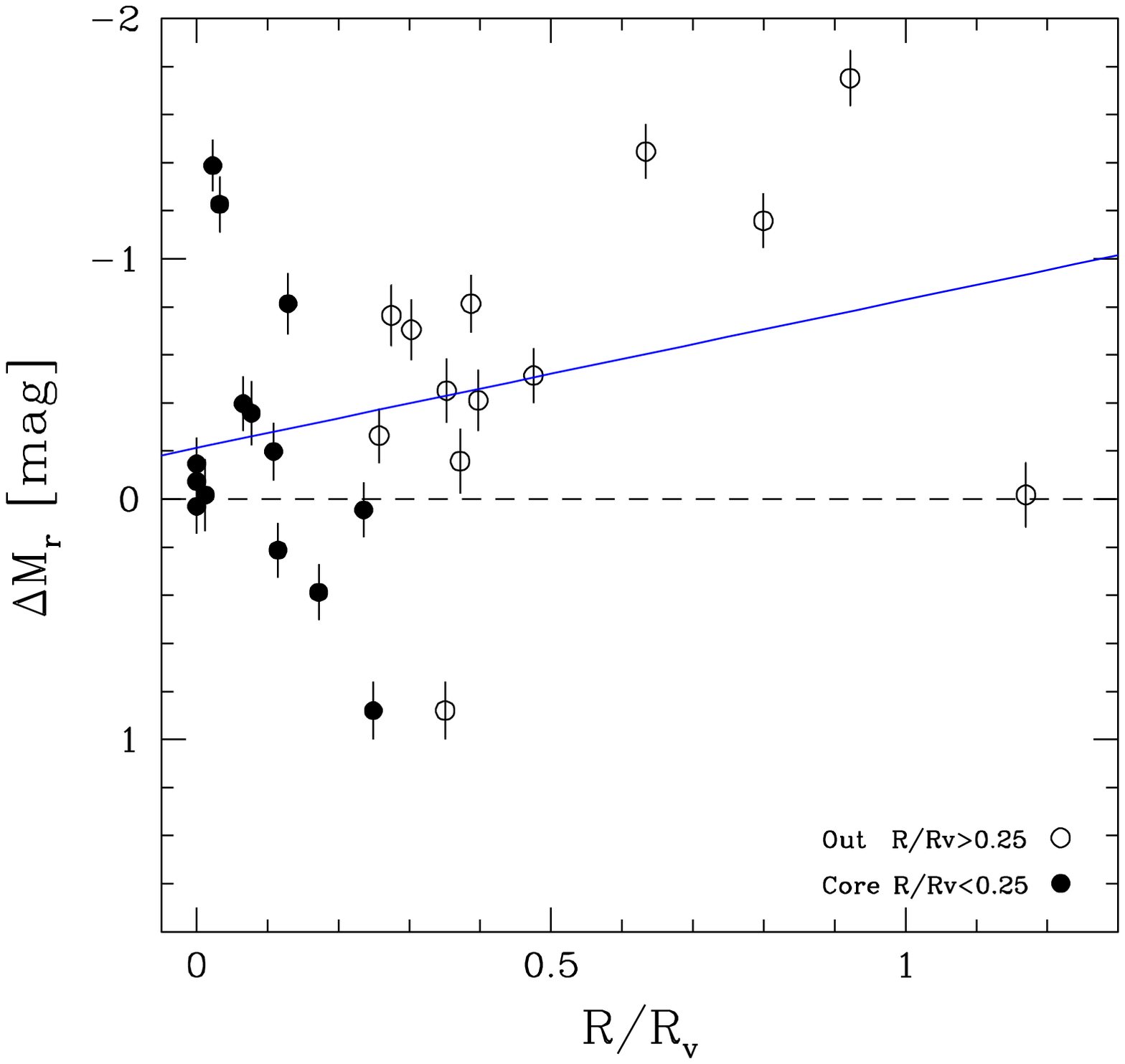}
\includegraphics[width=0.24\textwidth]{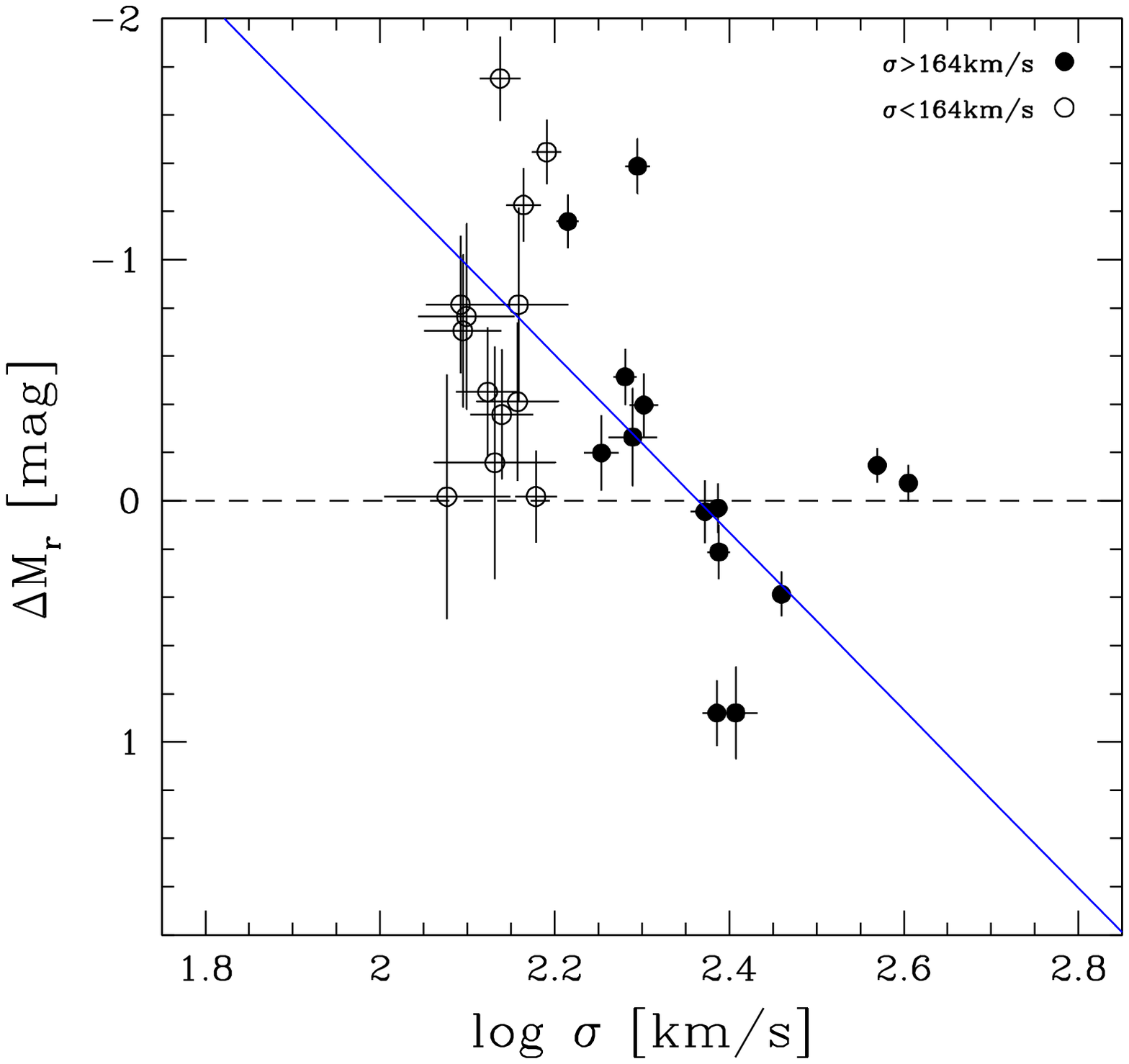}
\caption{\label{Fig1} Cluster Spheroidal Galaxies. Top row: rich clusters A\,2218 and A\,2390.
bottom row: low-mass clusters Cl\,0849, Cl\,1701 and Cl\,1702. Left panel:
Offsets from the FJR as a function of clustercentric radius. Right panel: FJR
offsets against velocity dispersion, which is a good indicator of the baryonic
mass. More massive galaxies and those residing in the cluster core are indicated
as solid symbols, less-massive ones and galaxies in the outer regions are shown
as open symbols (Fritz et al. 2005; Fritz 2006).}
\end{figure}

Field spheroidal candidates were selected based upon the deep $UBgRI$ and $BRI$
multi-band sky surveys of the FORS Deep Field (FDF, Heidt et al. 2003)
and William Herschel Deep Field (WHDF, Metcalfe et al. 2001), respectively.
For the follow-up VLT/FORS spectroscopy, the main selection criter\-ium was an
apparent brightness of $R\le22.0$ mag. This ensured a $S/N$$\ge$15 
$\AA^{-1}_{\rm rest}$ in the
absorption lines, mandatory for a robust measurement of internal kinematics 
and line-strengths. The final sample comprises 24 field sphe\-roidals down
to $M_B\le-19.30$, cf Fritz et al. (2009).

HST/WFPC2 and ACS structural parameters for 78 sph\-eroidal
galaxies were derived with two different algorithms, which analyse 
each galaxy surface brightness profile with seeing-convolved,
sky-corrected $r^{1/4}$ and exponential components, both
simultaneously and separately.
Elliptical (E) and S0 galaxies were morphologically classified
using a visual inspection, the profile fitting results
and their bulge-to-total fractions. A weak correlation
of the bulge fraction of cluster galaxies with visual morphological Hubble type
was found. Velocity dispersions ($\sigma$) based on  
$\langle S/N\rangle$$\ge$30$\AA^{-1}$ spectra could be derived
for 110 E+S0s. To ensure a reliable error treatment, 
Monte Carlo simulations for different stellar templates were performed 
(Fritz 2006; Fritz et al. 2009).

\section{Results}

\subsection{Cluster spheroidal Galaxies}

Scaling relations of the Faber--Jackson (FJR) and Kormendy relation as well as
the Fundamental Plane (FP) for distant spheroidal galaxies at $z\sim0.2$ obey
tight relationships similar to the local Universe. Assuming the local slope
holds valid with increasing redshift, on average a modest luminosity 
evolution of the galaxies is derived, regardless of their environment.
With respect to the local Coma cluster (J{\o}rgensen et al. 1995), spheroidal
galaxies in distant rich clusters display on average a moderate brightening of
$-0.35$ mag in the Gunn $r$-band. For spheroidal galaxies in three
low-mass clusters a median evolution of $\Delta M_{r}=-0.38$ mag is
derived. There is no significant increase of distant FJR scatter.
Compared to the local counterparts, the distant FJR have
shallower slopes at the 2$\sigma$ significance level, which implies
a \emph{mass-dependent luminosity evolution}.

Fig.~\ref{Fig1} shows the offsets of the FJR in Gunn $r$ for the
spheroidal galaxies in two rich (upper panel) and three low-mass clusters 
(lower panel) from the local FJR of the Coma sample as a function of
clustercentric radius and mass. 
Cluster spheroidals show a slight radial dependence. Galaxies
in the outer regions at $\ga$0.25 virial radii ($\ga$0.46 Mpc)
show a stronger brightening by $\sim$0.2\,mag than their central counterparts.
This radial trend with a stronger evolution for galaxies residing in the
outskirts of a cluster is also found in poor clusters, see Fig.~\ref{Fig1}.
Lower-mass galaxies in all cluster environments are on average brighter 
by $\sim$\,0.6\,mag in rest-frame Gunn~$r$ than locally, whereas the distant
spheroidals with masses of $M\approx 3\times 10^{11}$\,$M_{\sun}$ display only
a mild evolution with respect to their present-day counterparts.

Lenticular (S0) galaxies in rich clusters exhibit a stron\-ger evolution
compared to the local Coma FP than the Es. The stellar populations of
these systems cou\-ld be more diverse or comprise more complex star formation
histories (SFH; Fritz et al. 2005). This gets support by
the rest-frame galaxy colors. Bluer galaxies ($(B-I)<2.29$) show larger FJR
residuals from the local FJR of $\Delta M_{r}=-0.50$ mag, whereas redder
galaxies are less offset $\Delta M_{r}=-0.14$ mag. 
In both color ranges, S0 galaxies indicate a stronger evolution (brighter
by $\sim$0.7\,mag for the bluer and $\sim$0.45\,mag for the redder),
but blue Es have on average only a mild brightening of
$\sim$0.3\,mag than the local reference.

\subsection{Field spheroidal galaxies}

\begin{figure*}
\includegraphics[width=0.40\textwidth]{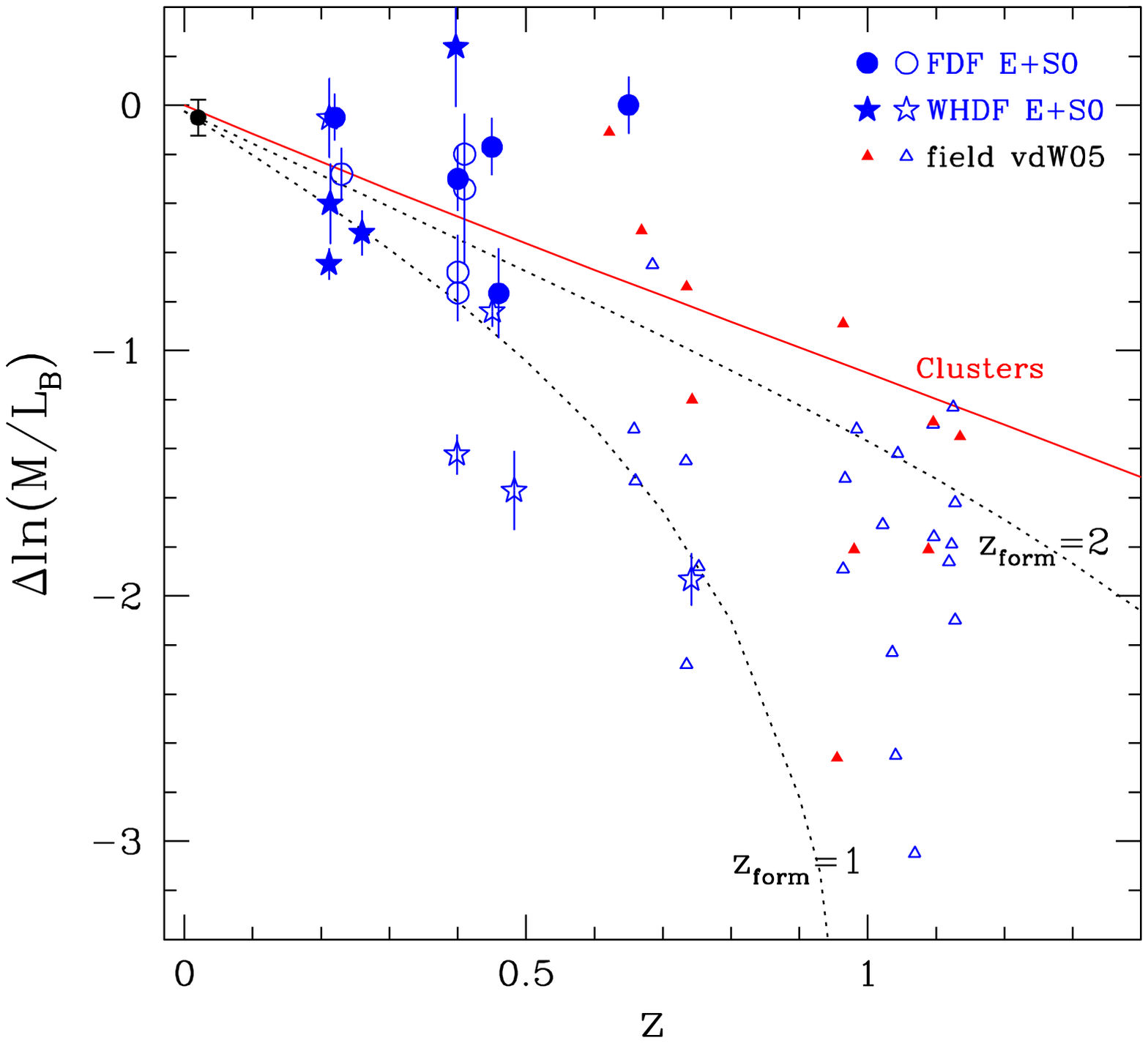}\hspace*{1cm}
\includegraphics[width=0.48\textwidth]{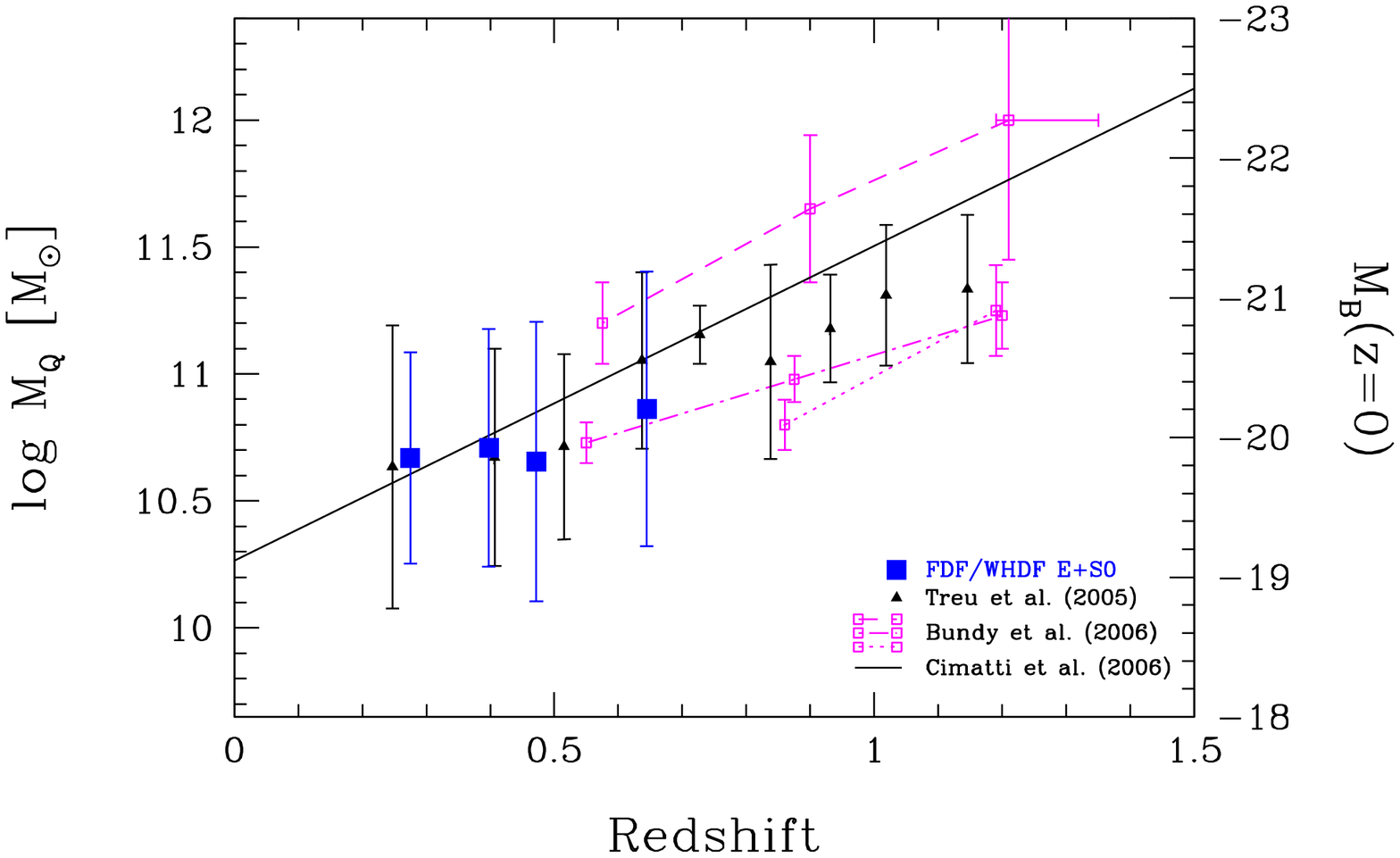}
\caption{\label{Fig2} Field Spheroidal Galaxies. Left: Evolution of the
mass-to-light ($M/L$) ratios. Less-massive galaxies ($M<2\times10^{11}M_{\sun}$)
indicate a faster evolution. Right: Evolution of the quenching mass. There is a
weak increase of the quenching mass with increasing redshift, in good
agreement with model predictions that suggest star formation triggering through
AGN feedback (Fritz et al. 2009).}
\end{figure*}

The mass-to-light ($M/L$) ratio of the field spheroidals evo\-lves on
average as $\Delta \ln({M/L_B})=-(1.71\pm0.18)\,z$, which is faster than for
cluster spheroidal galaxies. This is consistent with field spheroidals
comprising on average younger stellar populations than those of their cluster
counterparts. 

The evolution as derived from the scaling relations of the FJR, FP
and hence effective $M/L$ ratio shows a strong dependence on the mass.
Less-massive systems 
($M<2\times10^{11}M_{\sun}$) display a faster $M/L$ evolution 
$\Delta\nobreak\ln({M/L_B})=-(1.25\pm0.30)\,z$,
whereas more-massive galaxies ($M\nobreak>\nobreak2\times10^{11}M_{\sun}$) evolve 
slower $\Delta\nobreak\ln({M/L_B})\nobreak=\nobreak-(0.89\pm0.19)\,z$. This 
implies that for the most massive field galaxies the majority of their stellar
populations was formed early at $z_{\rm form}\nobreak=\nobreak3.5\pm1.3$.
For less-massive galaxies the $M/L$ evolution can be translated into 
$z_{\rm form}=\nobreak1.9\nobreak\pm\nobreak0.5$, which is at variance with the predictions of 
single-burst stellar population models. A fraction of 5\%--10\% in the total
stellar mass of lower-mass galaxies ($M<2\times10^{11}M_{\sun}$) must
have been formed at more recent epochs. At $z=0.4$, an accelerated evolution 
for less-massive galaxies ($M<2\times10^{11}M_{\sun}$) 
 ($\langle\Delta\nobreak\ln({M/L_B})\rangle\nobreak=\nobreak-0.54\pm0.07$)
than for more-massive ones ($\langle\Delta \ln({M/L_B})\rangle=-0.31\pm0.06$)
is significant on the 2$\sigma$ level (Fritz et al. 2009).

Field S0 galaxies feature a stronger luminosity evolution 
($\langle\Delta \ln({M/L_B})\rangle=-0.24\pm0.10$), 
bluer rest-frame colours and
more diverse stellar populations and hence evolve faster than field Es
($\langle\Delta \ln({M/L_B})\rangle=-0.09\pm0.07$). Interestingly, the
scatter appears to be mainly amplified for S0s.

Evidence for secondary star formation (SF) activity in the galaxies is
provided by \oii\,3727 emission or strong
H$\delta$ Balmer absorption as well as bluer rest-frame $(B-I)$ colour
diagnostics. The H$\delta$ absorption line strengths imply
that the residual (low level) SF of the galaxies accounts for
5\% to 10\% in the total stellar mass budget of these systems.

Over the past $\sim$6~Gyr, there is the trend of a slow
decreasing quenching mass (see Fig.~\ref{Fig2}), which characterises a crossover in
stellar mass above SF in galaxies gets suppressed. This suggests
that our systems experience a gradual suppression in their SF
processes. Our results favour a scenario where SF is not immediately suppressed
for less-massive halos and hence works on longer time scales in lower-mass
galaxies. For details, see Fritz et al. (2009).

\subsection{Conclusions}

The lack of an age difference found for field and cluster galaxies and the
dependence of the evolution on galaxy mass suggests that environmental effects
are not the dominant factors which drive the formation of early-type galaxies.
Internal properties, such as mass, size, $\sigma$ or chemical
composition, are the main contributors to their evolutionary history.
The evolution in the $M/L$ ratio in all environments favors a
\textit{down-sizing} formation scenario (Cowie et al. 1996).

A mass-dependent luminosity evolution is found in all environments
(Fritz et al. 2005; 2009). 
E and S0 galaxies are not a homogeneous group but
follow different evolutionary tracks. S0s indicate in all
environments a stronger evolution and more recent formation epochs
($1<z_{\rm form}\le2$). Thus S0s may have different stellar populations or
more complex, extended SFH, whereas Es evolve passively on
longer timescales yielding to older population ages
($z_{\rm form}\sim3$). 

\vspace*{0.04cm}

\acknowledgements
We thank the ESO Paranal and Calar Alto staff for efficient observational
support. This work has been supported by the VW Foundation (I/76\,520).
A.F. acknowledges partial support from grant HST-GO-10826.01 from STScI.
STScI is operated by AURA, Inc., under NASA contract NAS 5-26555.



\begin{thebibliography}{}

\bibitem[\protect\citeauthoryear{Baugh et al.}{1998}]{BCFL98}
Baugh, C.~M. et al.: 1998, ApJ, 498, 504

\bibitem[\protect\citeauthoryear{Cowie et al.}{1996}]{CSHC96}
Cowie, L.~L. et al.: 1996, AJ, 112, 839

\bibitem[\protect\citeauthoryear{De Lucia et al.}{2004}]{DLKW04}
De Lucia, G. et al.: 2004, MNRAS, 349, 1101

\bibitem[\protect\citeauthoryear{Fritz et al.}{2004}]{FZ04}
Fritz, A., \& Ziegler, B.~L.: 2004, in IAUC 195,
{\it Outskirts of Galaxy Clusters: Intense life in the suburbs},
ed. Diaferio, A., p. 502

\bibitem[\protect\citeauthoryear{Fritz et al.}{2005}]{FZBSD05}
Fritz, A. et al.: 2005, MNRAS, 358, 233

\bibitem[\protect\citeauthoryear{Fritz}{2006}]{F06}
Fritz, A.: 2006, PhD thesis, University of G\"ottingen

\bibitem[\protect\citeauthoryear{Fritz et al.}{2009}]{FBZ09}
Fritz, A., B\"ohm, A., \& Ziegler, B.~L.: 2009, MNRAS, 393, 1467

\bibitem[\protect\citeauthoryear{Heidt et al.}{2003}]{Hei03}
Heidt, J., et al.: 2003, A\&A, 398, 49

\bibitem[\protect\citeauthoryear{J{\o}rgensen et al.}{1995}]{JFK95b}
J{\o}rgensen, I. et al.: 1995, MNRAS, 273, 1097

\bibitem[\protect\citeauthoryear{Metcalfe et al.}{2001}]{Met01}
Metcalfe, N. et al.: 2001, MNRAS, 323, 779

\bibitem[\protect\citeauthoryear{Somerville \& Primack}{1999}]{SP99}
Somerville, R.~S., \& Primack, J.~R.: 1999, MNRAS, 310, 1087

\bibitem[\protect\citeauthoryear{Ziegler et al.}{2001}]{Z01}
Ziegler, B.~L. et al.: 2001, MNRAS, 325, 1571


\end{thebibliography}
\end{document}